\documentclass[twocolumn,trackchanges]{aastex701}
\usepackage{amsmath}
\usepackage{graphicx}
\usepackage{indentfirst}
\usepackage{float}
\usepackage{graphicx}
\usepackage{subcaption}
\usepackage{caption}
\usepackage{multirow}
\usepackage{CJK}
\usepackage{amsmath}  

\begin{document}
\begin{CJK*}{UTF8}{gbsn} 

\title{Extreme-Value Distribution Analysis of the Second CHIME/FRB Catalog: Assessing the Rarity of the One-off FRB 20250316A}

\correspondingauthor{Jun-Jie Wei}

\author[0009-0008-6247-0645]{Wen-Long Zhang (张文龙)}
\affil{Purple Mountain Observatory, Chinese Academy of Sciences, Nanjing 210023, China}
\affil{School of Astronomy and Space Sciences, University of Science and Technology of China, Hefei 230026, China}
\email{wlzhang@pmo.ac.cn}

\author[0000-0003-0162-2488]{Jun-Jie Wei (魏俊杰)}
\affil{Purple Mountain Observatory, Chinese Academy of Sciences, Nanjing 210023, China}
\affil{School of Astronomy and Space Sciences, University of Science and Technology of China, Hefei 230026, China}
\email[show]{jjwei@pmo.ac.cn}

\begin{abstract}
We present a statistical analysis of the extremely bright, apparently non-repeating fast radio burst FRB 20250316A, detected by the Canadian Hydrogen Intensity Mapping Experiment (CHIME), to assess its rarity.
Using a model-agnostic framework based on the Generalized Extreme Value (GEV) distribution and the second CHIME/FRB catalog, we perform Bayesian fits to the block-maxima of its peak flux and fluence. Our analysis confirms FRB 20250316A as a pronounced statistical outlier in both quantities. For the peak flux, the best-fit GEV model follows an unbounded, heavy-tailed Fr\'echet-type distribution, yielding return periods of approximately $802$ years at the $68\%$ confidence level (CL), $81$ years at the $95\%$ CL, and $30$ years at the $99\%$ CL.
The fluence distribution exhibits greater complexity: while the full sample is consistent with a Fr\'echet-type distribution (return period of approximately $55$, $15$, and $8$ years at the $68\%$, $95\%$ and $99\%$ CLs, respectively), removing three other conspicuous outliers reveals a light-tailed Weibull-type distribution with a finite upper bound that is far exceeded by the fluence of FRB 20250316A. Although its inferred recurrence time is shorter than that of the ``Brightest Of All Time'' (BOAT) gamma-ray burst GRB 221009A, FRB 20250316A represents a similarly exceptional event (a potential FRB ``BOAT'') within the relatively short observational baseline of wide-field radio surveys. This work affirms the existence of rare, extremely luminous events at the extreme upper end of the FRB luminosity distribution, which may delineate a distinct physical channel or the extreme tail of a complex luminosity function.
\end{abstract}


\section{INTRODUCTION} 
Fast Radio Bursts (FRBs) are millisecond-duration extragalactic radio transients that have posed a major astrophysical puzzle since the discovery of the first event (the ``Lorimer burst'') in 2007 \citep{2007Sci...318..777L,2019A&ARv..27....4P,2023RvMP...95c5005Z,2021SCPMA..6449501X,2025arXiv251019143K}. 

A wide range of theoretical models has been proposed to explain this enigmatic phenomenon. Advances in radio instrumentation and the growing body of observational data have, however, ruled out most of these proposals \citep{2019PhR...821....1P}. The leading contenders that remain are predominantly based on scenarios involving highly magnetized neutron stars (magnetars) capable of emitting coherent radio radiation. These include models of magnetars in binary systems with a companion star \citep{2020ApJ...893L..26I,2022NatCo..13.4382W,2025ApJ...994L..20Z} and the so-called ``belt and road'' model, in which a magnetar interacts with a surrounding asteroid belt \citep{2016ApJ...829...27D}. Although supported by mounting observational evidence \citep{2020ApJ...897L..40D,2020ApJ...898L..55G,2022NatCo..13.4382W,2023ApJ...942..102Z,2024ApJ...967L..44L,2025ApJ...994L..32L,2025arXiv251207140W,2025arXiv250517880Z}, such models generally predict repeating bursts. Observations, however, reveal that clearly repeating sources constitute only a small fraction of the known FRB population; most events appear non-repeating, with no repetitions detected to date. This dichotomy leads to two principal interpretations for apparently non-repeating FRBs: either their repetitions have not been observed due to instrumental limitations, or they originate from genuinely cataclysmic, one-off events.

Extensive studies of large samples have characterized the statistical properties \citep{2023ApJS..269...17H,2026MNRAS.545f2148N} and morphological classifications \citep{2021ApJ...923..230L,2022ApJ...939...27C,2023MNRAS.518.1629L,2023MNRAS.519.1823Z,2025ApJ...980..185S,2026arXiv260116048S,2025arXiv250906208K,2026MNRAS.545f2178M} of both repeating and non-repeating FRBs. Nevertheless, robust observational evidence distinguishing potential subclasses, particularly between genuine one-off events and as-yet-undetected repeaters, remains elusive.

A critical breakthrough occurred with the detection of FRB 20200428 from the Galactic magnetar SGR J1935+2154, which was spatially and temporally coincident with a bright X-ray burst \citep{2020Natur.587...54C,2020Natur.587...59B,2020ApJ...899L..27M,2020ApJ...898L..29M,2021NatAs...5..372R,2021NatAs...5..378L,2021ApJ...907L..17Z}. This association provided compelling evidence linking at least some FRBs to magnetar activity. Subsequent studies of this active episode further revealed distinctive properties of the accompanying X-ray bursts \citep{2023RAA....23k5013Z,2024ApJ...967..108X}.

While FRB 20200428 confirmed that magnetars can produce FRBs, it also highlighted unanswered questions about the diversity and extremes of such emissions. The event was relatively faint compared to the cosmological FRB population, leaving open the question of whether the same mechanism can explain the most luminous events observed at great distances. Investigating extreme outliers in luminosity is therefore crucial for testing the limits of magnetar models and for understanding the full population.

In this context, the Canadian Hydrogen Intensity Mapping Experiment (CHIME) detected an exceptionally bright burst, FRB 20250316A, 
on 16 March 2025. It exhibited a peak flux of $\mathrm{1.2\pm0.1\,kJy}$ and a fluence of $\mathrm{1.7\pm0.2\,kJy\,ms}$ \citep{2025ApJ...989L..48C}. Follow-up observations revealed no repetition, classifying it as a non-repeating event, and identified its probable host as the nearby spiral galaxy NGC 4141 \citep{2025ATel17081....1N,2025ATel17114....1A,2025ATel17124....1O}. 
However, deep multi-wavelength follow-up failed to detect any definitive X-ray, optical, or persistent radio counterparts, instead establishing stringent upper limits on such emission
\citep{2025ATel17112....1D,2025ATel17119....1S,2025ApJ...995....8L,2025ApJ...991L..20A,2025arXiv250904563K}. Thus, FRB 20250316A presents a contrasting case: an extremely luminous, apparently non-repeating event of extragalactic origin, with no detected multi-wavelength counterpart. Its exceptional brightness makes it a prime candidate for examing whether it represents a statistical extreme within the known FRB population, potentially challenging or refining models developed in the wake of the Galactic magnetar association.

\begin{figure}[http!]
\centering
\includegraphics[width=\columnwidth]{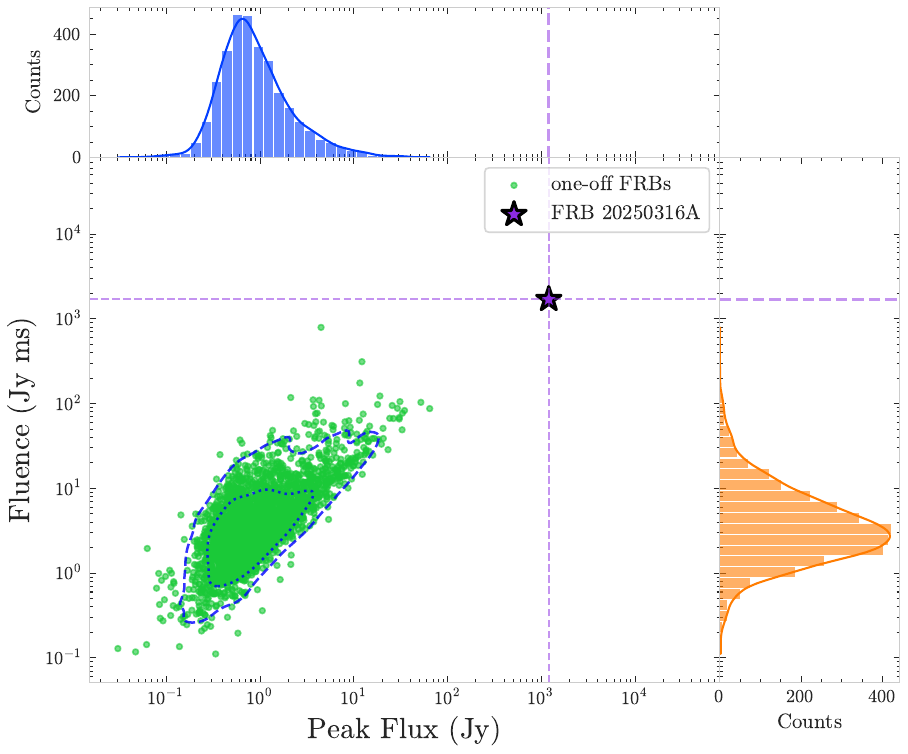}\\
\caption{Peak flux--fluence diagram for 3558 apparent non-repeaters from the second CHIME/FRB catalog (green points). The blue contour lines (from inner to outer) show the 68\% and 95\% credible regions of the joint distribution, as estimated via two-dimensional kernel density estimation.
The position of the extreme FRB 20250316A is marked by a star.} 
\label{flux_vs_fluence}
\end{figure}

The extraordinary luminosity of FRB 20250316A (see Figure~\ref{flux_vs_fluence}) marks it as a potential extreme outlier. To quantify its rarity, we take inspiration from the analysis of \cite{2025A&A...701A.109C}, who used the model-agnostic generalized extreme value (GEV) distribution to characterize the ``Brightest Of All Time'' (BOAT) gamma-ray burst GRB 221009A \citep{2023ApJ...946L..31B}. We apply a comparable GEV analysis to the recently released second CHIME/FRB catalog \citep{2026arXiv260109399F} to assess whether FRB 20250316A represents a similar statistical extreme---a potential `BOAT' among FRBs.

In this paper, we present an investigation into the extreme properties of FRB 20250316A through a GEV analysis of peak flux and fluence data from the catalog of 3558 non-repeating CHIME/FRB bursts. Section~\ref{sec:GEV} outlines the theoretical background of the GEV distribution and our Monte Carlo Markov Chain (MCMC) fitting methodology. Section~\ref{sec:results} presents and discusses the results, including the fitted GEV parameters and the derived recurrence probabilities. We conclude in Section~\ref{sec:conclusions} with a summary of our findings and their implications.

\section{Generalized Extreme Value Distribution Analysis}
\label{sec:GEV}
The GEV distribution provides a model-agnostic statistical framework for analyzing extreme values in observational data. Its principal strength lines in its asymptotic nature, which renders it largely independent of the underlying physical processes \citep{coles2001introduction}. Extreme value theory, which aims to quantify the occurrence probability of extreme measurements, is a rapidly evolving field in modern statistics and finds application across diverse disciplines including climatology, engineering, physics, and economics \citep{coles2001introduction}. In astronomy, recent implementations of this framework encompass the modelling of the brightest galaxies and active galactic nuclei \citep{2024MNRAS.534..173H, 2025MNRAS.543.3783H}, and the brightest GRB \citep{2025A&A...701A.109C}, as well as analyses of sunspot numbers \citep{2024JApA...45...14Z, 2025SoPh..300...50A}, cosmological density fields \citep{2018MNRAS.473.3598R}, galaxy cluster masses \citep{2012MNRAS.422.3554W}, and weak lensing shear peak counts \citep{2016MNRAS.456..641R}, among others.

\subsection{Data Reduction}
The GEV analysis is applied to block maxima. To implement this, the observational timeline is first partitioned into contiguous intervals of fixed duration. The maximum recorded value (e.g., peak flux or fluence) within each interval constitutes a block maximum, forming the dataset to which the GEV distribution is fitted. From the fitted parameters $(\mu,\,\sigma,\,\xi)$, one can calculate return levels and their confidence levels (CLs). The return level $z_{1/p}$ for a return period $1/p$ is defined as the value expected to be exceeded once, on average, over $1/p$ blocks, where each block contains $k$ observations. Uncertainties for all derived quantities are propagated from the posterior distribution of the GEV parameters.

The choice of block length $k$ involves a well-known bias-variance trade-off, as there is no first-principles method to determine an optimal value. The GEV distribution is an asymptotic limit; longer blocks (with more events per block) provide a better approximation to this limit but yield fewer block maxima, increasing the variance of parameter estimates. Conversely, shorter blocks provide more maxima for fitting but risk introducing bias if the block size is too small to approach the asymptotic regime. Therefore, the selection is typically iterative and data-dependent.

The second CHIME/FRB catalog spans approximately five years \citep{2026arXiv260109399F}. To balance the need for a sufficient number of blocks against the requirement for stable extreme value selection within each block, we adopt a block length of 30 days (see panel (a) of Figures~\ref{flux}, \ref{fluence}, and \ref{fluence_3_max}).

To assess the consistency of empirical return level curves derived from different block sizes, we compute the range-normalized root mean square error (NRMSE) between each candidate curve and a reference curve obtained with a 30-day blocking interval. For two sets of return levels $\{y_i^{\rm ref}\}$ (reference) and $\{y_i\}$ (candidate) evaluated at a common set of return periods, the NRMSE is defined as
\begin{equation}
    \mathrm{NRMSE} = \frac{\sqrt{\frac{1}{M} \sum_{i=1}^{M} \left( y_i - y_i^{\rm ref} \right)^2}}{y_{\rm max}^{\rm ref} - y_{\rm min}^{\rm ref}},
\end{equation}
where $M$ is the number of evaluation points, and $y_{\rm max}^{\rm ref}$ and $y_{\rm min}^{\rm ref}$ are the maximum and minimum values of the reference return level curve, respectively. Because the denominator depends only on the reference curve, this normalization renders the metric dimensionless and facilitates comparison across different scaling regimes. Following established practice in extreme value analysis and geophysical modeling \citep{coles2001introduction,willmott2005advantages,chai2014root}, 
we adopt a threshold of $\mathrm{NRMSE} < 0.15$ as a criterion for practical consistency between two return level curves. Values below this threshold indicate that the differences between curves are small relative to the overall dynamic range of the reference return level curve, which in turn implies robustness to the choice of block size.

\begin{figure}[http!]
\centering
\includegraphics[width=\columnwidth]{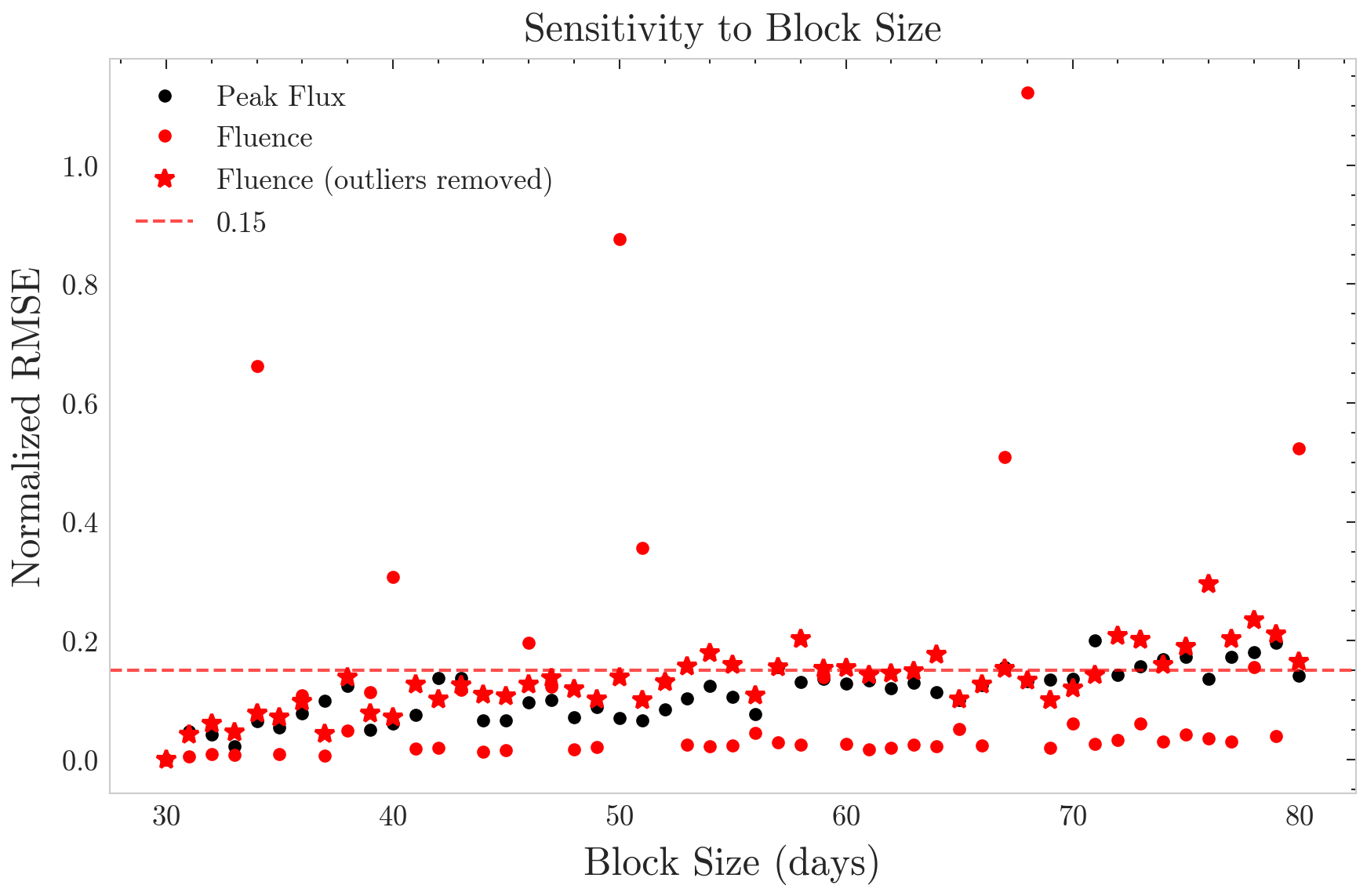}\\
\caption{NRMSE of empirical return level curves (computed relative to the 30-day block size reference) 
as a function of block duration. The dashed horizontal line denotes the $\mathrm{NRMSE}=0.15$ threshold, adopted as 
a criterion for practical consistency. The generally low NRMSE values across most block sizes support the robustness 
of our chosen 30-day blocking scheme.} 
\label{nrmse}
\end{figure}

The evolution of NRMSE as a function of block size is shown in Figure~\ref{nrmse}. For the peak flux, the NRMSE remains uniformly low across all block sizes (black dots), confirming that the 30-day reference curve is stable with respect to block size choice. The fluence generally exhibits similarly low NRMSE values (red dots), with most falling well below the 0.15 threshold, indicating good agreement comparable to that of the peak flux. However, a few isolated block sizes yield anomalously high NRMSE, occasionally exceeding unity, which signals strong sensitivity of the fluence estimates at those specific scales. This anomalous behavior arises when the boundaries of a given blocking interval align with one or more extreme fluence events. Such alignment causes these extreme events to be captured as block maxima, thereby disproportionately affecting the empirical tail of the distribution and inflating the NRMSE. Supporting this interpretation, when the three highest fluence values are excluded from the dataset, the anomalously high NRMSE spikes disappear entirely, and the metric remains consistently low across all block sizes (see red stars). These observations confirm that the occasional instability is not intrinsic to the underlying process but rather an artifact of rare, high-amplitude outliers interacting with the blocking scheme, a conclusion further supported by subsequent modeling analyses.

\subsection{Distribution Function}
For a sequence of independent and identically distributed random variables, the distribution of block maxima converges asymptotically to the GEV distribution. Its cumulative distribution function is
\begin{equation}
G(z) = \Pr\left(Z_{\mathrm{max}, n} \leq z\right) = \exp\left\{-\left[1 + \xi \left( \frac{z - \mu}{\sigma} \right) \right]^{-1/\xi}\right\},
\end{equation}
defined for all \(z\) satisfying \(1 + \xi (z - \mu)/\sigma > 0\). Here, \(\mu \in \mathbb{R}\) denotes the location parameter, \(\sigma > 0\) the scale parameter, and \(\xi \in \mathbb{R}\) the shape parameter. The shape parameter $\xi$ governs the tail behavior and categorizes the extreme-value distribution.

Specifically, the value of $\xi$ signifies the existence of an upper bound. A value of $\xi > 0$ corresponds to the Fr\'echet-type distribution, which originates from unbounded, heavy-tailed parent distributions and implies no theoretical upper limit for the extremes. In the limit \(\xi \to 0\), the distribution reduces to the Gumbel type,
\begin{equation}
G(z) = \exp\left\{-\exp\left(-\frac{z - \mu}{\sigma}\right)\right\}, \quad z \in \mathbb{R},
\end{equation}
which describes light-tailed distributions where the probability of exceeding a high threshold decays exponentially. Conversely, \(\xi < 0\) corresponds to the Weibull-type distribution, indicating a light-tailed parent distribution with a finite upper bound given by \(\mu - \sigma/\xi\).

\subsection{Fitting Procedure}
We divide the observational timeline into contiguous 30-day blocks. For each block, we extract the maxima (i.e., the highest peak flux and the highest fluence) to create separate datasets ${z_i}$ for each quantity. These block-maxima datasets are then fitted to the GEV distribution within a Bayesian framework. We sample the posterior distribution using the affine-invariant MCMC ensemble sampler implemented in the $emcee$\footnote{https://pypi.org/project/emcee/} package for Python.

The log-likelihood function $\ell(\mu,\,\sigma,\,\xi \mid {z_i})$ accounts for the support of the distribution and the Gumbel limit ($\xi \to 0$). For a block maximum $z_i$, we define $s_i = (z_i - \mu)/\sigma$. The total log-likelihood $\ell = \ln L(\mu,\,\sigma,\,\xi \mid {z_i})$ is the sum over all data points, where the likelihood $L$ given by:
\begin{align}
\nonumber &L(\mu,\sigma,\xi\mid {z_i}) = \prod_{i=1}^n g(z_i) \\
&= \begin{cases}
\displaystyle\prod_{i=1}^n \frac{1}{\sigma} e^{-s_i - e^{-s_i}}, & |\xi| \to 0, \\
\displaystyle\prod_{i=1}^n \frac{1}{\sigma} (1+\xi s_i)^{-(1+\frac{1}{\xi})} e^{-(1+\xi s_i)^{-1/\xi}}, & \xi \neq 0,
\end{cases}
\end{align}
with the condition $1 + \xi s_i > 0$ required for all $i$ when $\xi \neq 0$. Here, $g(z)$ is the probability density function of the GEV distribution, $g(z) = dG(z)/dz$.

To avoid prior-driven biases and ensure that the inference is dominated by the data, we adopt uniform priors for the location (\(\mu\)), scale (\(\sigma\)), and shape (\(\xi\)) parameters. This choice is justified by \cite{northrop2016posterior}, since our sample size is far larger than the requirement minimum of four. These priors are sufficiently flexible to capture Fr\'echet-type (\(\xi>0\)), Gumbeltype (\(\xi\to0\)), or Weibull-type (\(\xi<0\)) tail behavior, as informed by the interpretation above.

After obtaining the posterior samples, we calculate return levels. The $T$-year return level $z_T$, defined as the value expected to be exceeded once per $T$ years on average, is obtained by inverting the cumulative distribution function: $G(z_T) = 1 - 1/T$. 
Within the context of this fitted GEV model, an event whose observed value significantly exceeds the return levels associated with high quantiles (e.g., those at the 68\% or 95\% CLs) can be considered a statistical outlier, suggesting it represents an exceptionally rare realization of the underlying population described by the catalog data.

\section{Results and Discussion}
\label{sec:results}

As shown in Figure~\ref{flux_vs_fluence}, FRB 20250316A is a distinct outlier in both peak flux and fluence, underscoring its extreme nature.

\subsection{Peak Flux Distribution}
\label{subsec:flux}
Figure~\ref{flux} displays the GEV fit to the peak flux distribution for 3558 apparently non-repeating FRBs from the second CHIME/FRB catalog.
Panel (a) shows the time series of 30-day block maxima. 
No significant secular trend is observed, supporting the stationarity assumption for the extreme-value process. The peak flux of FRB 20250316A markedly exceeds all other block maxima, visually confirming its outlier status.

Panel (b) presents the fitted GEV model, which quantifies the theoretical relationship between
peak flux and return period. For FRB 20250316A ($\mathrm{1.2\pm0.1\,kJy}$), the model yields return periods of $802$ years at the 68$\%$ CL, $81$ years at the 95$\%$ CL, and $30$ years at the 99$\%$ CL. The posterior distribution of the shape parameter $\xi$ (panel c) has a mean of $0.29 > 0$, indicating that the underlying extremal distribution is of the the unbounded, heavy-tailed Fr\'echet type. Finally, panel (d) assesses the goodness-of-fit using a quantile-quantile (QQ) plot. The data points largely fall within the $\sim$2$\sigma$ CL, indicating a reasonable fit consistent with expectations given the limited sample size.

\subsection{Fluence Distribution}
\label{subsec:fluence}
Figure~\ref{fluence} presents the GEV analysis of the fluence distribution for apparently non-repeating FRBs. 
Panel (a) shows FRB 20250316A as a clear outlier among the 30-day block maxima.
A closer inspection of the return-level curve (panel b) reveals a secondary feature: three other bursts (FRBs 20200723B, 20220222B, and 20210922C, denoted by red points) deviate significantly from the trend defined by the bulk of the data. Their positions suggest a leverage effect on the GEV fit, which may bias the tail estimation. Including these three outliers could substantially influence both the inferred shape parameter and the return period for FRB 20250316A, thus affecting our assessment of its rarity.

To rigorously evaluate the influence of these points and the robustness of our conclusions, we perform a comparative analysis. Instead of removing only the target source, we consider two scenarios: one using the full sample (including the three outliers) and another excluding them. This allows us to assess the sensitivity of the results to the presence of these outliers. We first consider the complete sample (Figure~\ref{fluence}). Within this framework, the model assigns return periods of 55 years (68\% CL), 15 years (95\% CL), and 8 years (99\% CL) to the fluence of FRB 20250316A (panel b). The estimated shape parameter is $\xi = 0.35 > 0$ (panel c), consistent with a heavy-tailed, unbounded Fr\'echet-type distribution. However, the goodness-of-fit, evaluated via the QQ plot (panel d), is poor, with substantial deviations beyond the 95\% credible band. This strongly suggests that the fit to the complete sample is unduly influenced by the three ancillary outliers.

\begin{figure*}[http!]
\centering
\includegraphics[width=0.5\textwidth, angle=0]{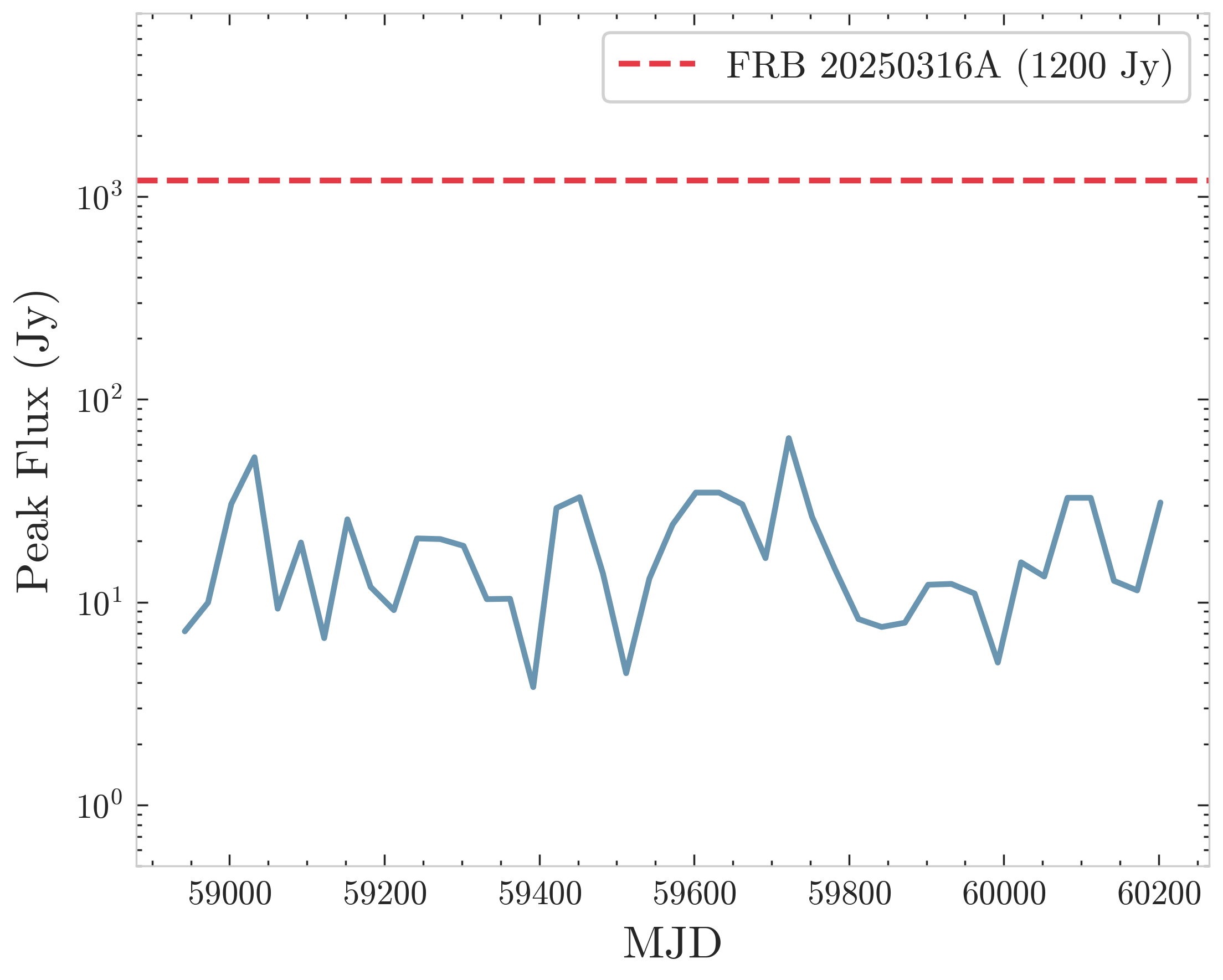}\put(-240, 210){\bf (a)}
\includegraphics[width=0.5\textwidth, angle=0]{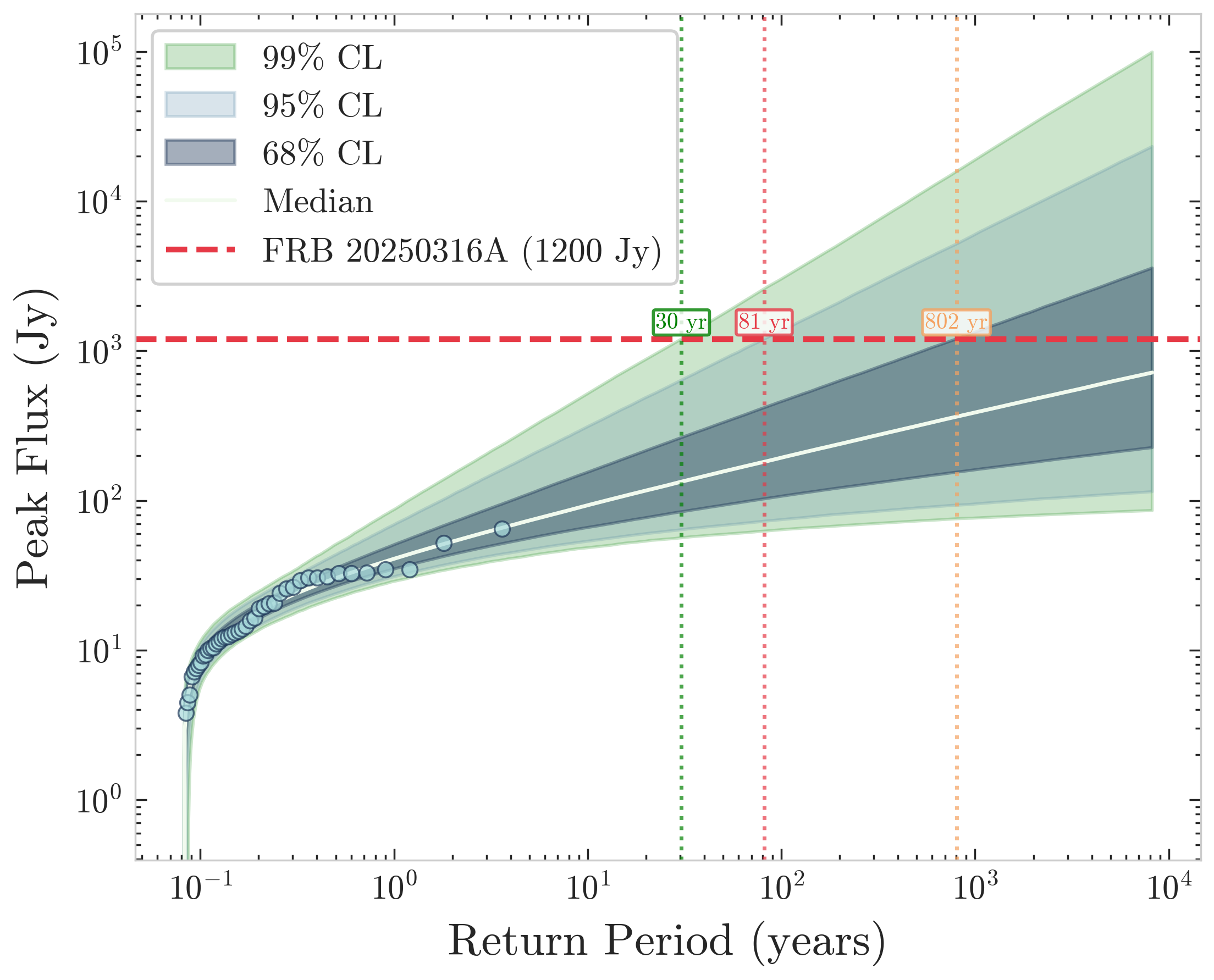}\put(-240, 210){\bf (b)}\\
\includegraphics[width=0.5\textwidth, angle=0]{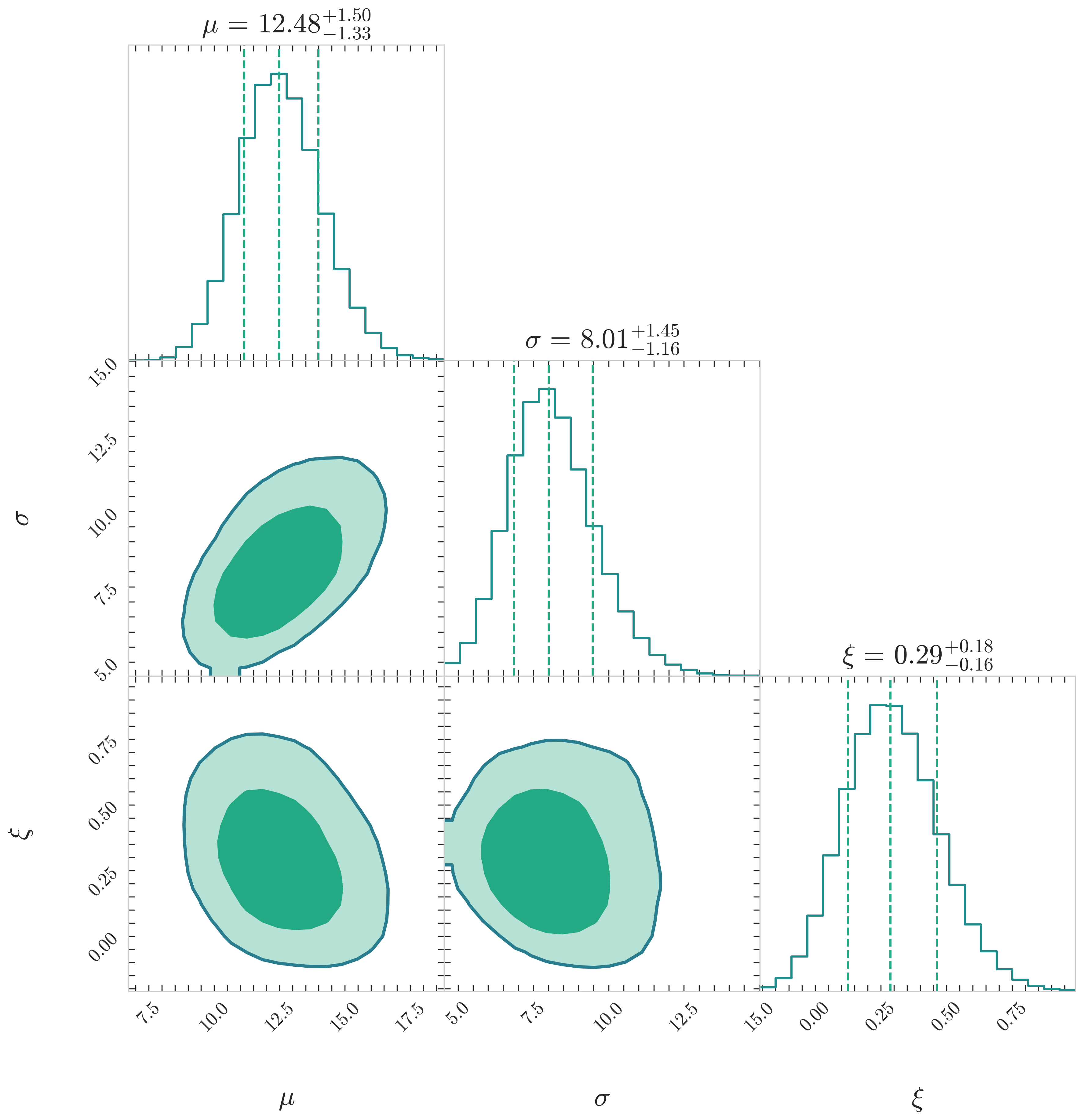}\put(-240, 210){\bf (c)}
\includegraphics[width=0.5\textwidth, angle=0]{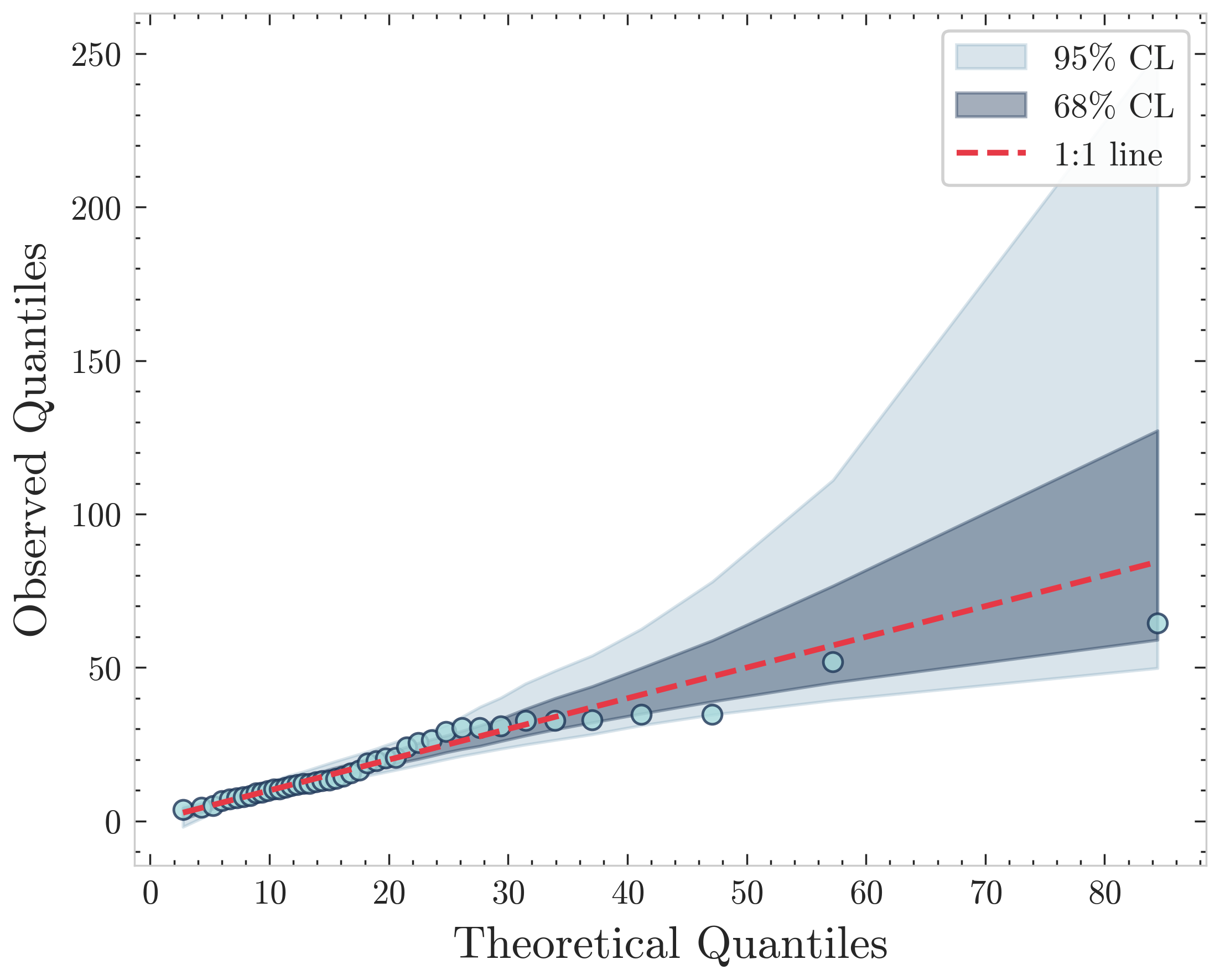}\put(-240, 210){\bf (d)}\\
\caption{GEV analysis of the peak flux distribution for 3558 apparently non-repeating FRBs from the second CHIME/FRB catalog. \textbf{(a)} Time series of the 30-day block maxima. The red dashed line marks the peak flux of the extreme outlier FRB 20250316A. \textbf{(b)} Return plots for GEV fit of extreme values (solid curve). The intersection of this curve with the flux of FRB 20250316A (horizontal dashed line) corresponds to return periods of 802, 81, and 30 years at the 68\%, 95\% and 99\% CLs, respectively. \textbf{(c)} 1D posterior distributions and 2D credible regions (showing the 68$\%$ and 95$\%$ contours) for the GEV parameters ($\mu$, $\sigma$, and $\xi$). \textbf{(d)} Quantile-Quantile (Q-Q) plot for assessing the goodness-of-fit between the empirical block maxima and the fitted GEV distribution. The 68\% and 95\% credible regions are also shown.}
\label{flux}
\end{figure*}

\begin{figure*}[http!]
\centering
\includegraphics[width=0.5\textwidth, angle=0]{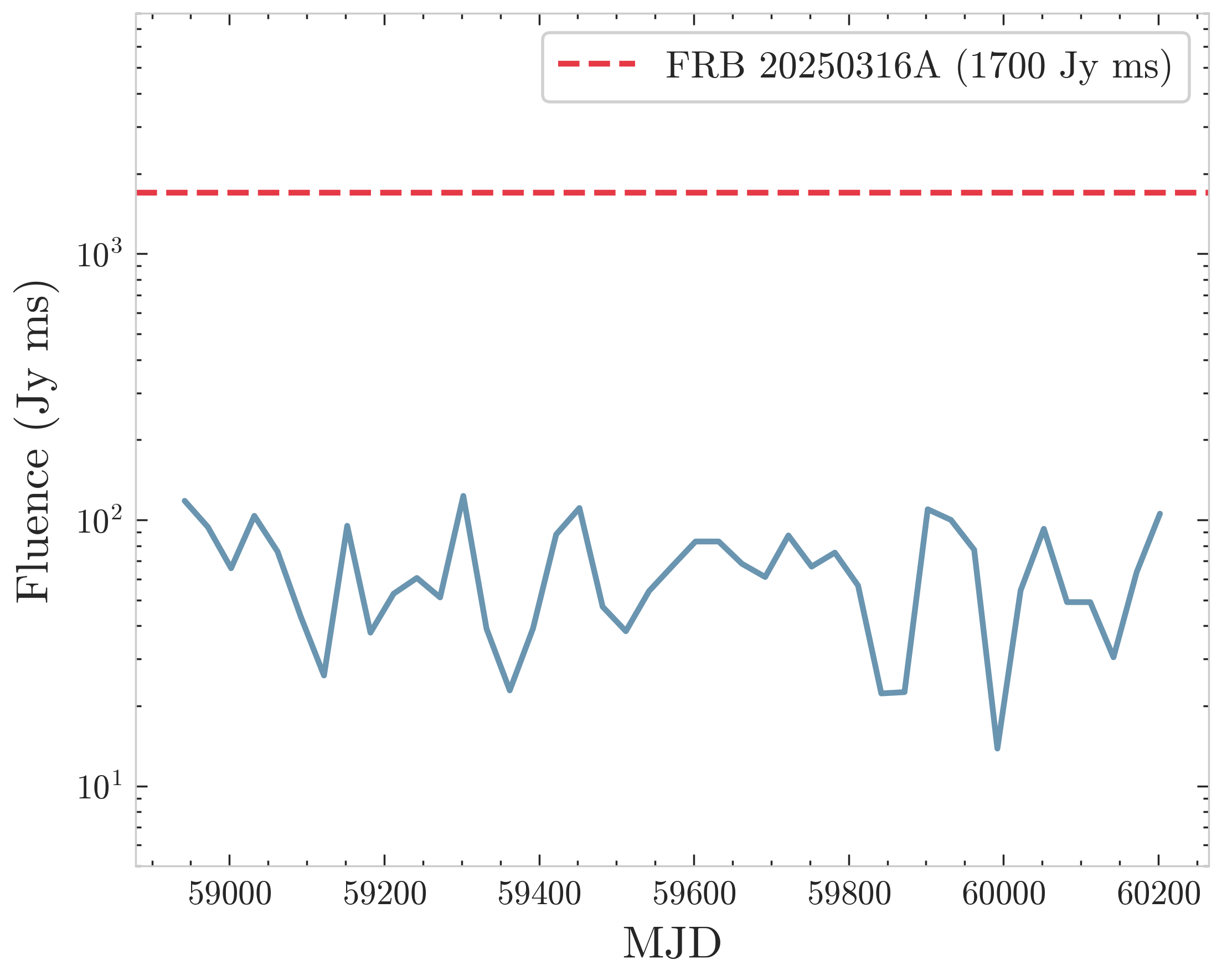}\put(-240, 210){\bf (a)}
\includegraphics[width=0.5\textwidth, angle=0]{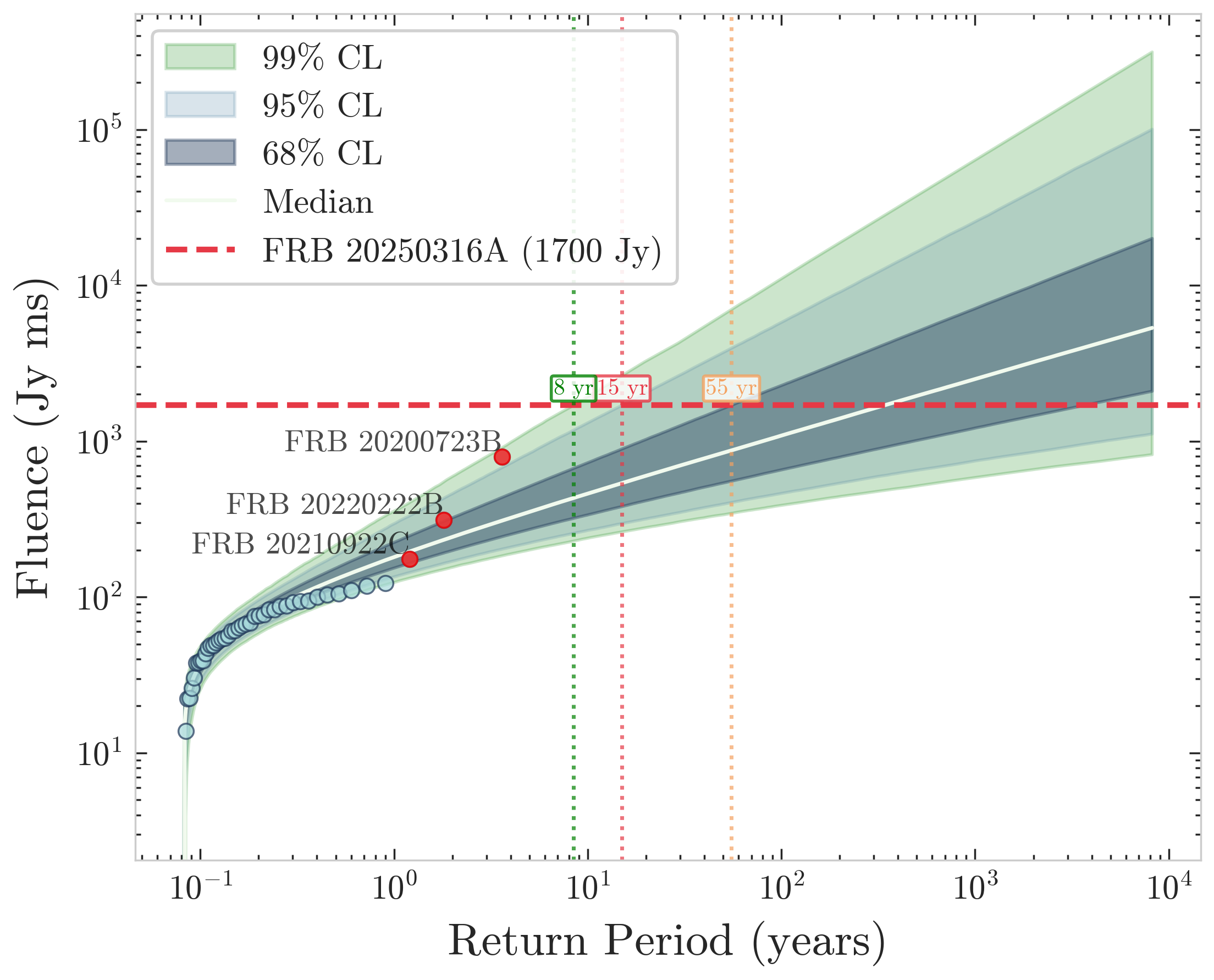}\put(-240, 210){\bf (b)}\\
\includegraphics[width=0.5\textwidth, angle=0]{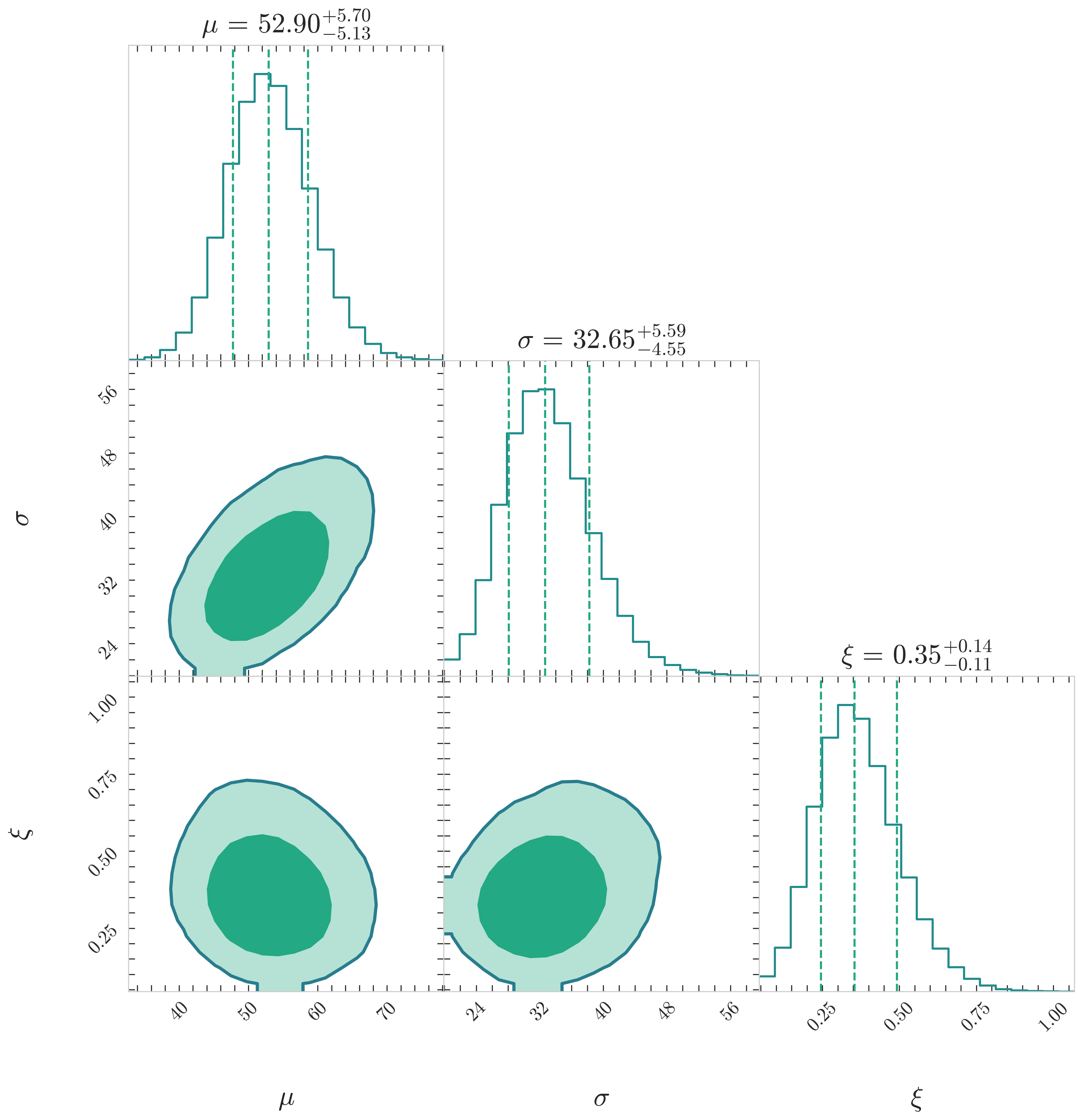}\put(-240, 210){\bf (c)}
\includegraphics[width=0.5\textwidth, angle=0]{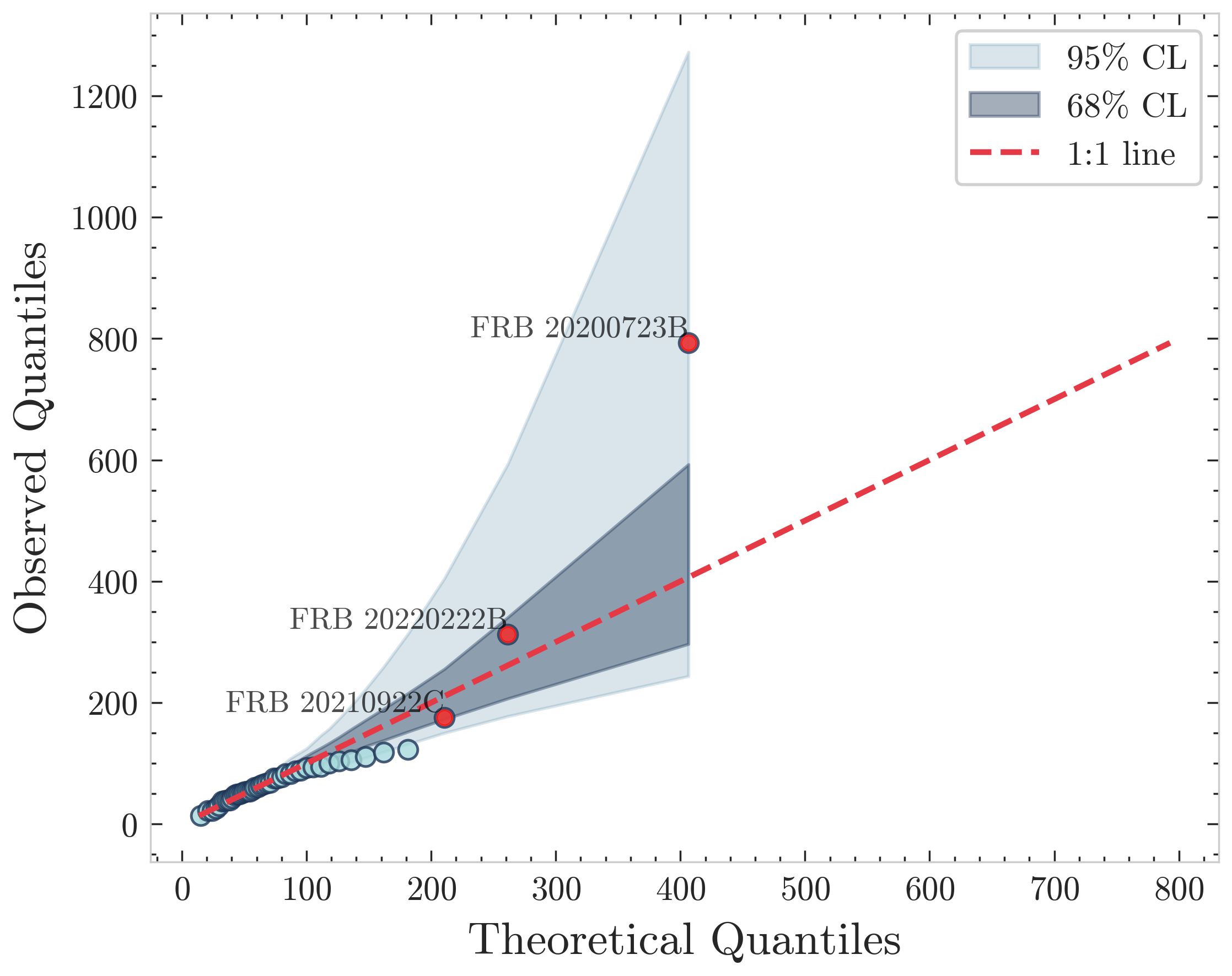}\put(-240, 210){\bf (d)}\\
\caption{Same as Figure~\ref{flux}, but showing the GEV analysis of the fluence distribution for 3558 apparently non-repeating FRBs.
Three bursts (FRBs 20200723B, 20220222B, and 20210922C) that deviate from the bulk trend are denoted by red symbols.}
\label{fluence}
\end{figure*}

\begin{figure*}[http!]
\centering
\includegraphics[width=0.5\textwidth, angle=0]{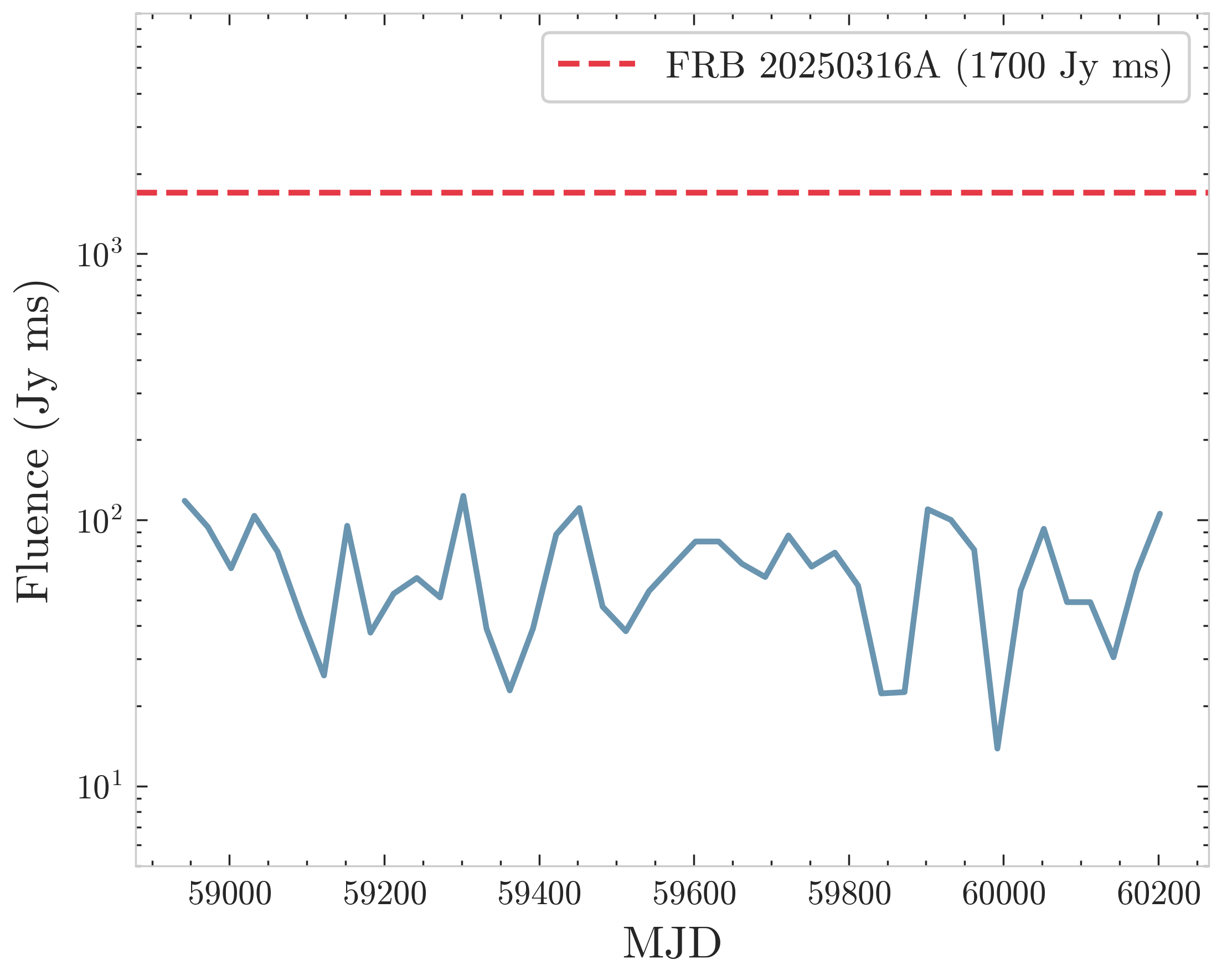}\put(-240, 210){\bf (a)}
\includegraphics[width=0.5\textwidth, angle=0]{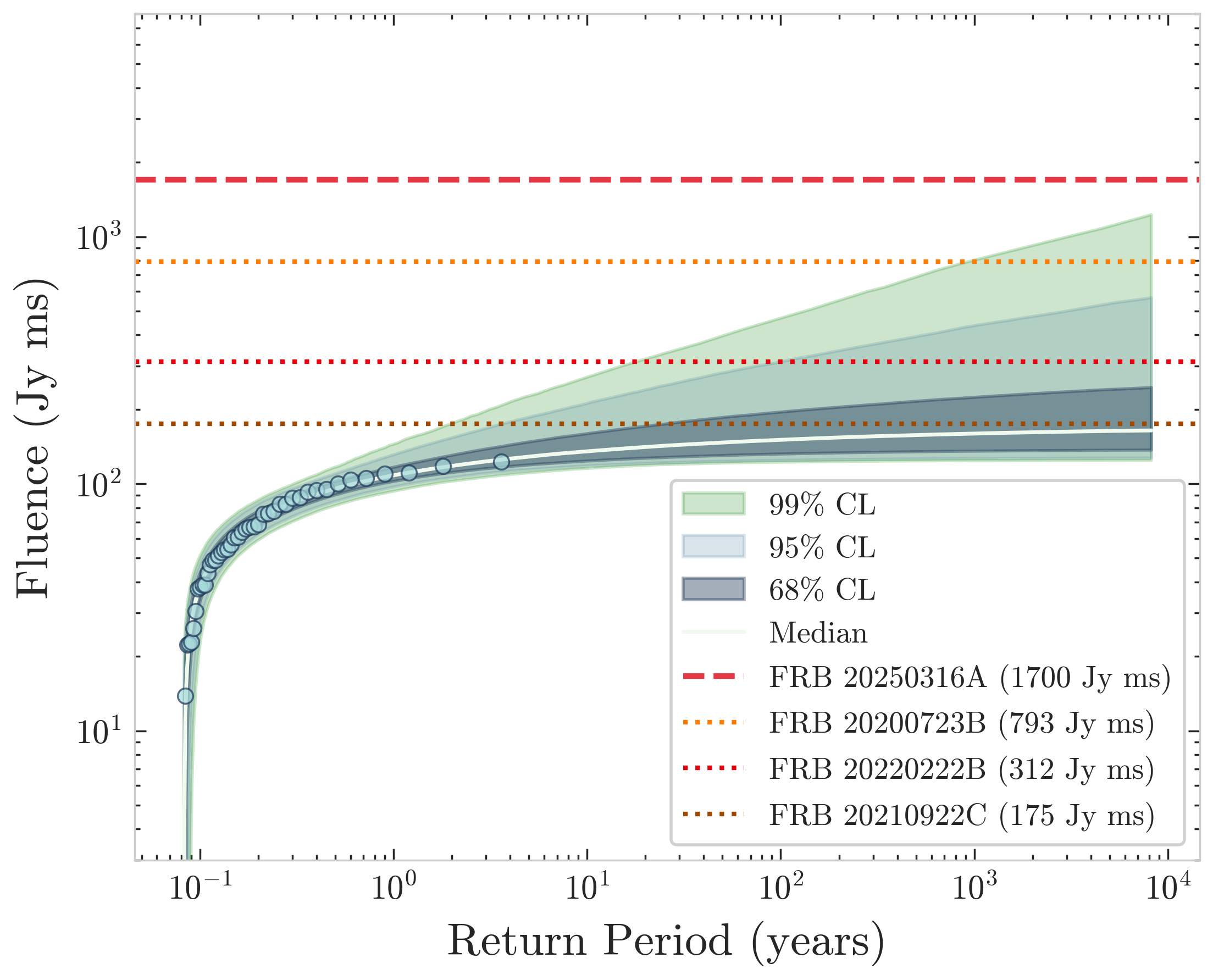}\put(-240, 210){\bf (b)}\\
\includegraphics[width=0.5\textwidth, angle=0]{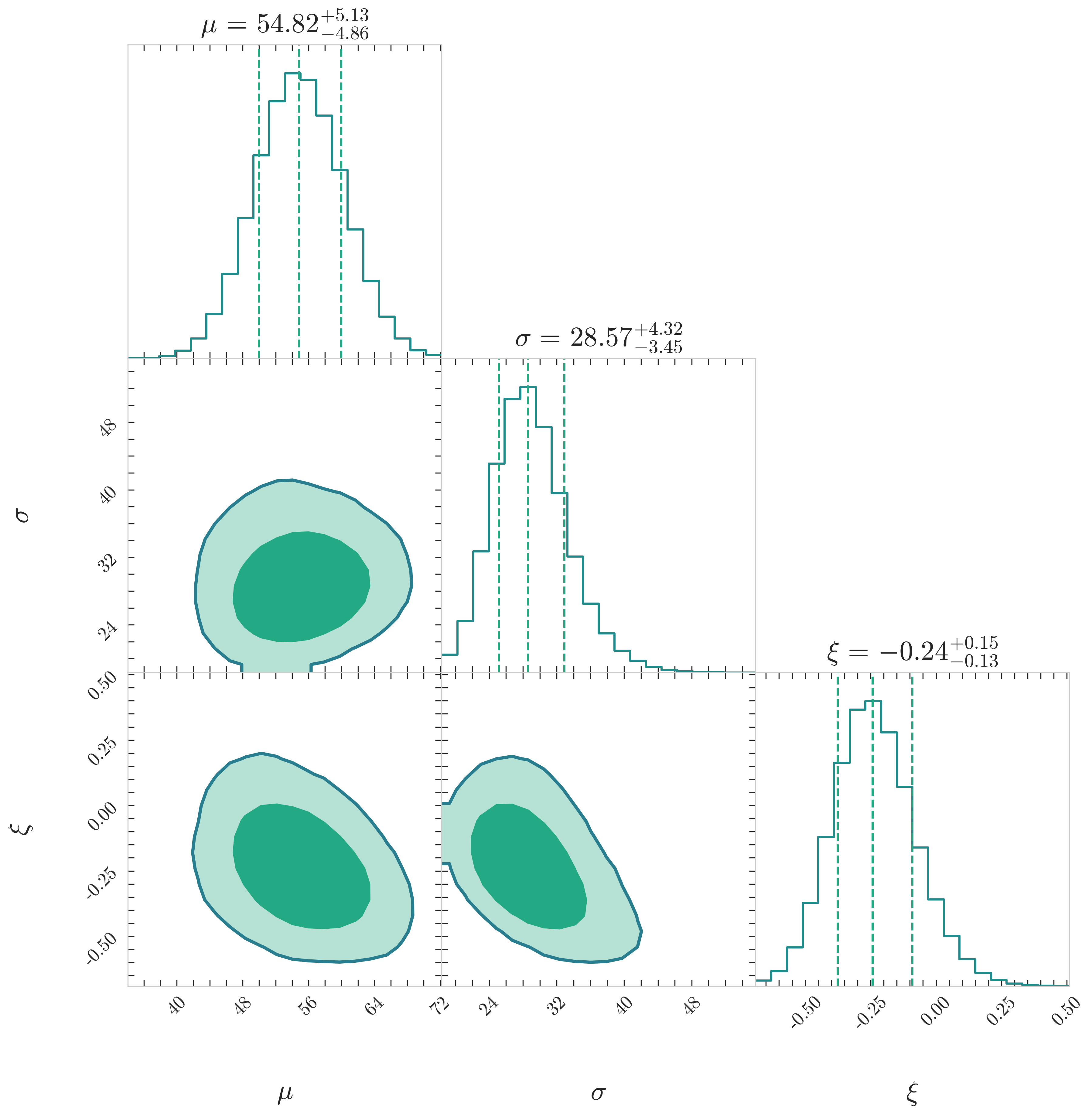}\put(-240, 210){\bf (c)}
\includegraphics[width=0.5\textwidth, angle=0]{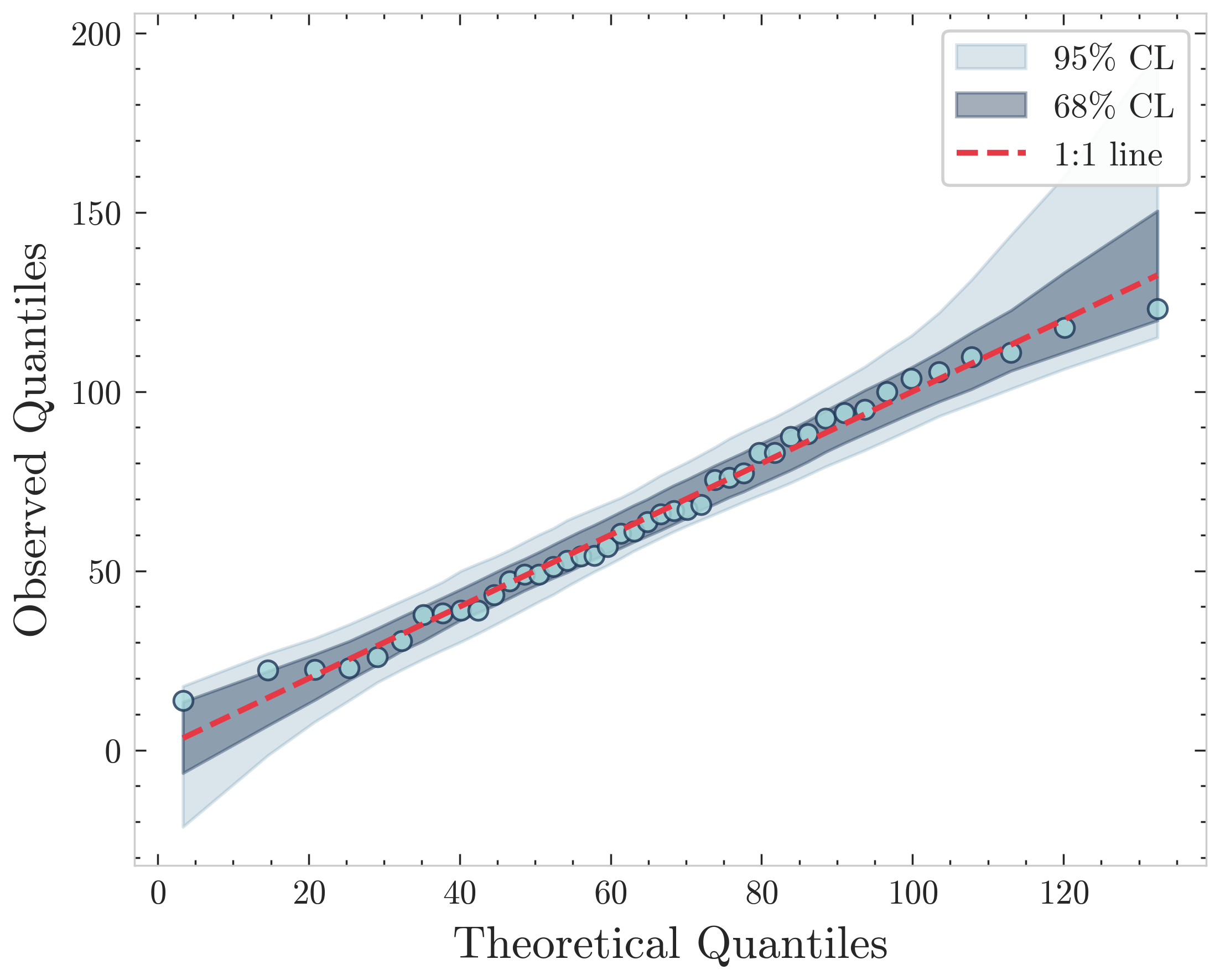}\put(-240, 210){\bf (d)}\\
\caption{Corresponding GEV analysis for the fluence distribution after removing three extreme outliers (FRBs 20200723B, 20220222B, and 20210922C). The sample now contains 3555 apparently non-repeating FRBs.}
\label{fluence_3_max}
\end{figure*}

Since outliers may represent a distinct phenomenological class \citep{2021MNRAS.508...69K}, we exclude FRBs 20200723B, 20220222B, and 20210922C to analyze the core population in isolation (Figure~\ref{fluence_3_max}). This censoring leads to a fundamental shift in the inferred extremal behavior. The posterior mean of the shape parameter becomes $\xi = -0.24 < 0$ (panel c), characterizing the distribution as a light-tailed Weibull-type with a finite upper bound of $\mu - \sigma/\xi \approx 173$ Jy ms. The fluence of FRB 20250316A ($\mathrm{1.7\pm0.2\,kJy\,ms}$) vastly exceeds this bound. This scenario---an extreme event lying orders of magnitude beyond the estimated upper limit of its putative parent population---qualitatively mirrors the statistical narrative of GRB 221009A \citep{2025A&A...701A.109C}. Importantly, the posterior for $\xi$ retains support for positive values, meaning a Fr\'echet-type distribution remains plausible, albeit at lower confidence. Finally, as panel (b) shows, even within this refined context where the three excluded FRBs are themselves rare, FRB 20250316A maintains its status as a far more extreme statistical outlier. In contrast, for the peak flux distribution (Figure~\ref{flux}), no such conspicuous outliers (apart from FRB 20250316A itself) are present; the block maxima follow a relatively clean trend. Therefore, removing additional data points in the peak flux analysis was neither necessary nor justified.

\section{Conclusions}
\label{sec:conclusions}

In this study, we have applied a model-agnostic statistical framework based on the GEV distribution, together with the second CHIME/FRB catalog of non-repeating bursts, to quantitatively assess the rarity of the extremely luminous FRB 20250316A.

Our analysis shows that FRB 20250316A is a pronounced statistical outlier in both peak flux and fluence. For the peak flux, the best-fit GEV distribution is of the Fr\'echet-type distribution (shape parameter $\xi > 0$), yielding return periods of $802$ years at the 68\% CL, $81$ years at the 95\% CL, and $30$ years at the 99\% CL.

The fluence distribution presents a more complicated picture owing to three other additional outliers. When these events are retained, the collective sample favors a heavy-tailed Fr\'echet-type distribution, in which the fluence of FRB 20250316A corresponds to return periods of 55 years (68\% CL), 15 years (95\% CL), and 8 years (99\% CL). If, instead, these three outliers are treated as a distinct subclass and removed, the remaining core population is better described by a light-tailed Weibull-type distribution ($\xi < 0$) with a finite upper bound $\mu - \sigma/\xi \approx 173$ Jy ms. The fluence of FRB 20250316A ($\mathrm{1.7\pm0.2\,kJy\,ms}$) lies far beyond this bound. This behavior---an extreme event exceeding the estimated upper limit of its putative parent distribution---statistically parallels that of the extreme transient GRB 221009A. We note that in this case the posterior distribution of the shape parameter $\xi$ spans both negative and positive values, reflecting the inherent uncertainty in tail characterization when only a limited number of extreme events is available.

Our results firmly establish FRB 20250316A as one of the most extreme non-repeating FRBs yet detected. Its properties challenge models that treat non-repeating FRBs as a single homogeneous population, suggesting instead that the brightest FRBs may arise from distinct physical channels or represent the rare tail of a complex luminosity function. Although the inferred return periods are shorter than the multi-millennial scale estimated for GRB 221009A, a characteristic recurrence time of decades to centuries remains statistically extraordinary given the relatively short operational baseline of current wide-field radio surveys. Future, larger samples from CHIME/FRB and other upcoming instruments will be essential to refine these statistical estimates, probe the true shape of the FRB luminosity function, and ultimately uncover the physical origin of such extreme cosmic explosions.

\begin{acknowledgments}
We are grateful to the anonymous referee for helpful comments.
We would like to acknowledge helpful discussions with Yi-Fang Liang and Wen-Jun Tan, and the use of DeepSeek for language polishing.
This work is supported by the National Natural Science Foundation of China (grant Nos. 12422307, 12373053, and 12321003)
and the CAS Project for Young Scientists in Basic Research (grant No. YSBR-063).
\end{acknowledgments}


\bibliography{sample701}{}
\bibliographystyle{aasjournalv7}

\end{CJK*}
\end{document}